\documentclass[12pt,letterpaper]{article}
\usepackage[utf8]{inputenc}

\pdfoutput=1
\usepackage{color}
\definecolor{darkblue}{rgb}{0.1,0.1,.7}
\usepackage[colorlinks, linkcolor=darkblue, citecolor=darkblue, urlcolor=darkblue, linktocpage]{hyperref}
\usepackage[]{amsmath}
\usepackage[]{graphicx}
\usepackage[]{latexsym}
\usepackage[utf8]{inputenc}
\usepackage{slashed,graphicx,color,amsmath,amssymb}
\usepackage[mathscr]{eucal}
\usepackage{mathrsfs}
\usepackage[margin=10pt,font=small,labelfont=bf]{caption}
\usepackage{amscd}
\usepackage{bm}
\usepackage{xcolor}
\usepackage{upgreek}
\usepackage[square, comma, sort&compress,numbers]{natbib}
\usepackage[all,cmtip]{xy}
\usepackage[margin=1.5in]{geometry}
\usepackage{cleveref}
\usepackage[color=cyan!30!white,linecolor=red,textsize=footnotesize]{todonotes}
\usepackage{microtype}
\usepackage{soul}
\numberwithin{equation}{section}
\numberwithin{figure}{section}

\hypersetup{
	pdfstartview={FitH},    
	pdftitle={Higher spin wormholes from modular bootstrap},    
	pdfauthor={Das, Datta},     
	colorlinks=true,       
	linkcolor=blue,          
	citecolor=blue!59!black! ,        
	filecolor=magenta,      
	urlcolor=blue           
}

\newcommand{\tr}{\mathrm{Tr}\,}

\def\btau{{\bar{\tau}}}
\def\Z{\mathbb{Z}}
\def\pd{\partial}

\def\E{{\mathcal{E}}}

\def\ie{{\it i.e.~}}
\def\eg{{\it e.g.~}}

\def\nn{\nonumber}
\def\pd{\partial}

\def\l1{{{1-loop}}}

\def\im{{{Im}}}

\def\n1{\Bigg|_{n=1}}

\def\n{{(n)}}
\def\tr{{Tr}}

\def\tr{\text{Tr}}

\def\E{\mathcal{E}}

\def\pd{\partial}

\def\beq{\begin{equation}}
\def\eeq{\end{equation}}

\def\bea{\begin{eqnarray}}
\def\eea{\end{eqnarray}}
\def\nn{\nonumber}
\def\pd{\partial}

\def\l1{{\text{1-loop}}}

\def\im{{\text{Im}}}

\def\n1{\Bigg|_{n=1}}

\def\n{{(n)}}
\def\tr{\text{Tr}}

\def\ket#1{|#1\rangle}

\def\vev#1{\langle{#1}\rangle}

\def\be{\begin{equation}}
\def\ee{\end{equation}}
\def\bal{\begin{array}{l}}
\def\ba#1{\begin{array}{#1}}  
	\def\ea{\end{array}}
\def\bea{\begin{eqnarray}}
\def\eea{\end{eqnarray}}
\def\beas{\begin{eqnarray*}}
	\def\eeas{\end{eqnarray*}}

\def\nn{\\\nonumber}

\def\vev#1{\langle #1 \rangle}

\def\nn{\nonumber}
\def\bit{\begin{item}}
	\def\eit{\end{item}}
\def\benu{\begin{enumerate}}
	\def\eenu{\end{enumerate}}
\def\tr{{\rm tr}}

\def\cW{\mathcal{W}}

\DeclareFontFamily{U}{wncy}{}
\DeclareFontShape{U}{wncy}{m}{n}{<->wncyr10}{}
\DeclareSymbolFont{mcy}{U}{wncy}{m}{n}
\DeclareMathSymbol{\Sha}{\mathord}{mcy}{"58}

 \makeatletter
 \g@addto@macro\bfseries{\boldmath}
 \makeatother

\makeatletter
\if@todonotes@disabled

\else

\fi
\makeatother

\begin{document}

\definecolor{tinge}{RGB}{255, 244, 195}
\sethlcolor{tinge}
\setstcolor{red}

\vspace*{-.8in} \thispagestyle{empty}
\begin{flushright}
	\texttt{CERN-TH-2021-085}
\end{flushright}
\vspace{.7in} {\Large
\begin{center}
{\Large \bf  Higher spin wormholes from modular bootstrap}
\end{center}}
\vspace{.4in}
\begin{center}
{Diptarka Das$^1$ \& Shouvik Datta$^2$}
\\
\vspace{.4in}
\small{
	$^1$  \textit{Department of Physics, Indian Institute of Technology - Kanpur,\\
		Kanpur 208016, India.}\\
	\vspace{.5cm}
  $^2$\textit{Department of Theoretical Physics, CERN,\\
	1 Esplanade des Particules, Geneva 23, CH-1211, Switzerland.}\\
\vspace{.5cm}

} \vspace{0cm}
\begingroup\ttfamily\small
didas@iitk.ac.in,~sdatta@cern.ch\par
\endgroup


\end{center}

\vspace{.6in}

\begin{abstract}
\normalsize
We investigate the connection between spacetime wormholes and ensemble averaging in the context of higher spin AdS$_3$/CFT$_2$. 
Using techniques from modular bootstrap combined with some holographic inputs, we evaluate the partition function of a Euclidean wormhole in AdS$_3$ higher spin gravity. The fixed spin sectors of the dual CFT$_2$  exhibit features that starkly go beyond conventional random matrix ensembles: power-law ramps in the spectral form factor and potentials with a double-well/crest underlying the level statistics. 
\end{abstract}

\vskip 0.7cm \hspace{0.7cm}

\newpage

\setcounter{page}{1}

\noindent\rule{\textwidth}{.1pt}\vspace{-1.2cm}
\begingroup
\hypersetup{linkcolor=black}
\tableofcontents
\endgroup
\noindent\rule{\textwidth}{.2pt}

\setlength{\abovedisplayskip}{14pt}
\setlength{\belowdisplayskip}{14pt}

\section{Introduction}
\label{sec:intro}

The mechanism by which quantum information escapes an evaporating black hole is one of the deepest mysteries of modern theoretical physics. Over the past few years, there has been significant progress on this front that reproduce a unitary Page curve from semi-classical gravity path integrals \cite{Almheiri:2019qdq,Penington:2019kki,Almheiri:2020cfm}. A crucial ingredient in the analysis involves the inclusion of wormhole saddles that interpolate between regions connected by an entanglement cut. In a similar vein, Euclidean wormholes can also connect two separate  boundaries.
 In the context of AdS/CFT, the existence of such wormhole solutions lead to a loss of factorization in the observables of, what is apparently, a direct product of CFTs \cite{Marolf:2020xie,Maldacena:2004rf}. 

A promising way out of this conundrum is to formulate versions of holography in which the dual CFT isn't a single theory but an ensemble average of theories. This idea finds its origins in the context of spin-glasses where the effective description emerges from  a disorder-average over Hamiltonians. The partition function is then given by the mean of the partition functions of the ensemble, $\vev{Z(\beta)}$, while the non-vanishing fluctuations or higher moments, $\vev{Z(\beta_1)Z(\beta_2)\cdots}$, encode the wormhole amplitudes of the bulk dual. Such a construction is largely motivated by the fact that  topological JT gravity in 2d is precisely dual to an ensemble of random matrices \cite{SSS}. In one dimension higher, similar ideas have been developed which demonstrate that pure AdS$_3$ gravity shares common features with random matrix theory (RMT) and perturbative $U(1)^D\times U(1)^D$ Chern-Simons (CS) theory is dual to a theory of $D$ free bosons averaged over Narain moduli 
\cite{Afkhami-Jeddi:2020ezh,Maloney:2020nni,Perez:2020klz,Datta:2021ftn,Benjamin:2021wzr,Ashwinkumar:2021kav,Dong2021}.

This brings us to a natural question: how fundamental is the notion of ensemble averaged holography? Ultimately, one would like to depart from a semi-classical gravity approximation and understand whether ensemble averaging makes sense in  string theoretic constructions of AdS/CFT. String theory in AdS$_3$ with a single unit of NS-NS flux has been shown to be exactly dual to the symmetric orbifold of $\mathbb{T}^4$ \cite{Eberhardt:2019ywk}. In this setup, it has been demonstrated that wormhole partition functions do factorize and, therefore, the averaging operation is unnecessary at the free orbifold point \cite{Eberhardt:2020bgq}. We are then inclined to ask: where does the averaged description break down? 

As we lack fundamental principles at this point, it is valuable to explore whether ensemble averaging can be embedded into more general settings and see what lessons these situations can offer. In this work, we explore this possibility  in  the framework of  higher spin AdS$_3$/CFT$_2$. Theories of massless higher spin fields describe the leading Regge trajectory of string theory in the tensionless limit. These theories are grown-up versions of classical (super)gravity theories and are more tractable than full-fledged string theories. As the symmetries of the gravity theory get enhanced beyond diffeomorphisms, we lose traditional geometrical notions such as horizons and geodesics. 
The CFT duals, described by coset constructions, have additional  higher spin conserved currents that enlarge the chiral algebra to $\cW_\infty$ \cite{Gaberdiel:2010pz}. The coset models are however rational CFTs and do not possess the features of sparseness or chaos which are central to describe black hole microstates. We shall therefore focus on irrational CFTs with $\cW_N$ symmetries (with $c>N-1$) which are dual to finite tower higher spin fields in AdS$_3$, often dubbed pure higher spin gravity \cite{Alday:2020qkm}. 

The key object we consider in this paper is the partition function of a Euclidean wormhole in higher spin gravity. This wormhole connects two spacetime boundaries which are tori. As the precise details of the CFT dual are a priori unclear, we employ modular bootstrap techniques (along with some well-informed assumptions from holography) to arrive at the partition function. In particular, we adapt the techniques of \cite{Cotler:2020ugk} to the case where the CFT has $\cW_N$ symmetries instead of Virasoro.
The analysis in this work, therefore, provides a concrete realization of the wider applicability of the methods to bootstrap ensembles. The wormhole partition function takes the form of a Poincar\'e series along with some prefactors encoding contributions from zero-modes and $\cW_N$ descendants. 
We shall see that the wormhole amplitude   appropriately generalizes the pure gravity case and, at first sight, has a  form very similar to the Narain moduli average of $(N-1)$ bosons. The zero-modes and descendant contributions turn out to be the same as the Narain averaged case but the Poincar\'e series itself is slightly different. This leads to drastic differences in the spectral correlations. 

We dissect the wormhole amplitude further by Fourier transforming to  sectors of fixed spin. 
A useful quantity that captures  statistics of energy eigenvalues is the spectral form factor: $\vev{Z(\beta+it)Z(\beta -it)}$. This quantity can be obtained from the correctly projected wormhole amplitude upon analytic continuation. We find that at late times the spectral form factor of a fixed spin sector strikingly displays a power-law ramp $\sim t^{N-1}$, in contrast to the linear one for RMT or the pure gravity/Virasoro case (for $N=2$). Although we haven't tracked down the species of RMTs that mimic this behaviour, the faster ramp ties in well with earlier findings that the irrational $\cW_N$ CFTs violate the bound on chaos, under certain approximations \cite{Perlmutter:2016pkf}. In the non-chaotic $N \rightarrow \infty$ limit, where the theory is described by a 't Hooft limit of rational coset CFTs \cite{Gaberdiel:2010pz}, the ramp might be expected to show exponential behaviour similar to integrable fermion models \cite{Liao:2020lac,Winer:2020mdc}. 

The pair correlation functions of the spectral densities exhibit some novel properties for the higher spin case. We evaluate this quantity directly from the wormhole amplitude using an inverse Laplace transform and, also independently, using the method of resolvents.  For even $N$, we find long-range correlations between the eigenvalues. Whereas for odd $N$, the spectral correlations turn out to be short-ranged and they localize around delta-function singularities. If at all a random matrix description exists for this, the potentials for the eigenvalues should have some regimes of attraction -- we verify this in a toy example. 
These properties are markedly different from the pure gravity counterpart and the lower-dimensional case of JT gravity. It is undoubtedly imperative to ask whether two-dimensional higher spin gravity, described by a topological BF theory \cite{Gonzalez:2018enk,Alkalaev:2019xuv,Alkalaev:2020kut,Alkalaev:2021zda}, also has these properties in its spectrum. We do not address this question in this paper, hoping to return to it in the near future. 

This paper is organized as follows. In \S \ref{sec:partition-function} we evaluate the wormhole partition function in higher spin gravity using modular bootstrap. This section contains an outline of the bootstrap method and the ingredients of $\cW_N$ CFTs we need for the analysis. We study the spectral statistics of wormhole partition function in \S \ref{sec:spectral-studies} -- this constitutes finding the spectral form factor, the pair correlation function of the spectral density and the potentials describing the level statistics. We conclude in \S \ref{sec:discuss} and discuss some avenues for future research. The appendices contain some identities of Bessel functions, additional technical details and consistency checks.


\section{Partition function of the Euclidean wormhole}
\label{sec:partition-function}

\begin{figure}[t!]
	\centering
	\includegraphics[width=1\linewidth]{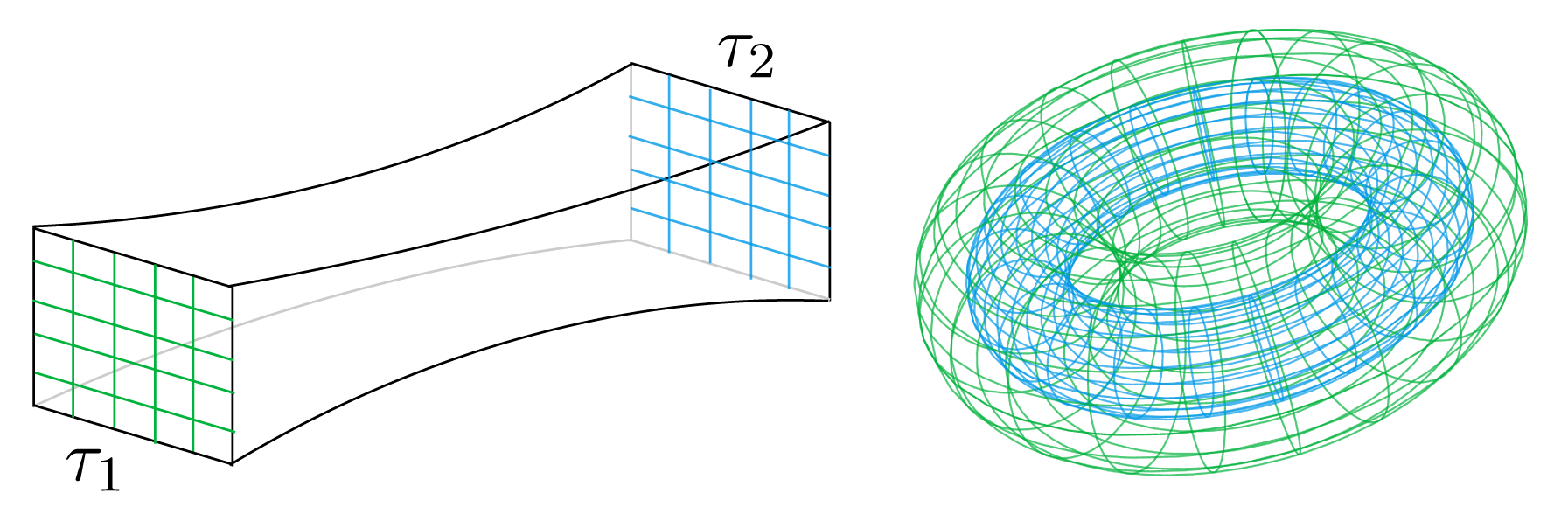}
	\caption{The topology of the Euclidean wormhole is that of torus$\times$interval, or equivalently annulus$\times$circle. The outer and inner tori boundaries have an opposite sense of orientation from the bulk point of view.}
	\label{fig:tori}
\end{figure}

In the context of AdS$_3$/CFT$_2$,  Euclidean wormholes have been studied for the pure gravity case in \cite{Cotler:2020hgz,Cotler:2020ugk}. 
The bulk topology of the 3d Euclidean wormhole is $\mathbb{T}^2\times I$, see Fig.\,\ref{fig:tori}. The two boundaries of the wormhole are given by two distinct tori, that are connected via the bulk geometry. 

The partition function of the wormhole (often referred to as the `wormhole amplitude') can be obtained from the gravitational path integral using a constrained instanton approach, and this method has been further systematized to higher dimensions \cite{Cotler:2020lxj,Cotler:2021cqa}.  In hindsight, it has been realized that the wormhole amplitude can be bootstrapped by imposing  modular constraints arising from the boundary tori. However, in this method, it is not just modular invariance alone that fixes the amplitude. Other essential inputs -- like smoothness, topological considerations, boundary orientation and charge conservation -- have bulk origins and play a key role in determining the partition function.

\subsection{The modular bootstrap procedure} \label{sec:procedure}

In this subsection we review the steps involved in the modular bootstrap procedure \cite{Cotler:2020hgz}. It can be seen from Fig.\,\ref{fig:tori} that the tori, living at the boundaries of $\mathbb{T}^2\times I$, have no relative Dehn twists and are oppositely orientated with respect to each other (\ie the outward normals point in opposite directions).
This feature implies that modular transformations act oppositely on the tori. We can define a double moduli \emph{preamplitude}, $\tilde{Z}\!\left( \tau_1, \tau_2 \right)$, that obeys the following invariance constraint:
\begin{align}
	\tilde Z\!\left( \tau_1, \tau_2 \right) &= \tilde Z\!\left( \gamma \tau_1 , \gamma^{-1} \tau_2 \right), \qquad  
	\gamma \in {\rm PSL}(2;\mathbb{Z}).\label{bootstrap1}
\end{align}
We have suppressed anti-holomorphic dependence to simplify the notation.
The action of $\gamma$ and $\gamma^{-1}$ denote the simultaneous modular and inverse modular transformations of $\tau_1$ and $\tau_2$
\begin{align}
{	\gamma \cdot \tau = \frac{a\tau+b}{c\tau+d }~, \quad ad - bc =1~. }
\end{align}
{For future reference, we note that the S-modular transformation is $\tau \mapsto -1/\tau$, while the T-modular transformation is $\tau\mapsto \tau+1$. These two transformations generate the modular group $SL(2,\mathbb{Z})$.}

Next, the preamplitude is proportional to the moduli space volume form, $V_0$, which arise in the bulk from the zero-mode contributions dictating the relative twist between the two tori. This is a physical effect, hence $\tilde{Z}$ is imbued with this contribution. Furthermore, as the two tori are the boundaries of the same connected bulk, charge conservation constrains the CFTs living on the two  tori boundaries to have the same spectrum of primaries. Requiring bulk-smoothness also keeps the conformal dimensions above the BTZ threshold. In the momentum representation, $h - \tfrac{c-c_{\text{curr}}}{24} = \frac{k^2}{4}$ (with $h$ representing the conformal dimension, $c$ the central charge and $c_{\text{curr}}$ the number of conserved currents), this implies that $0 \leq k \leq \infty$. With these inputs, one arrives at an useful ansatz for the preamplitude
\begin{align}\label{ansatz}
	\tilde{Z}(\tau_1,\tau_2) &= V_0 \int_0^\infty \,dk\, d\bar{k}\, \chi_k(\tau_1)\bar \chi_{\bar k}(\bar \tau_1)\chi_k(\tau_2)\bar \chi_{\bar k}(\bar \tau_2)\, \rho(k,\bar k),
\end{align}
where, $\chi_k(\tau)$ is the CFT character. The details of the character depend on the chiral algebra of the CFT which is also the asymptotic symmetry algebra of the bulk theory.   The bootstrap constraint \eqref{bootstrap1} is sufficient to determine the distribution of primaries, $\rho(k,\bar k)$, upto an overall normalization. Once this is obtained, we plug it back into \eqref{ansatz} to evaluate $\tilde{Z}(\tau_1, \tau_2)$. {Note that in the above ansatz we have implicitly assumed that the CFT in question is irrational, \ie we have an infinite number of primaries owing to modular invariance. Furthermore, the character appearing in \eqref{ansatz} will turn out to be non-degenerate characters of the chiral algebra which have no restrictions coming from null states.}

The quantity $\tilde{Z}(\tau_1,\tau_2)$, however, is not the full wormhole amplitude yet. It misses instances where only one of the two tori gets modular transformed. From the bulk point of view, these are distinct and allowed physical configurations. Therefore, the full partition function should involve a sum over them. Such configurations are generated by $\gamma$ acting on one of the torus moduli. We finally end up with
\begin{align}\label{mod-sum-0}
	Z(\tau_1, \tau_2) &= \sum_{\gamma\,\in\, {\rm PSL}(2,\mathbb{Z} )} \tilde{Z}( \tau_1, \gamma \tau_2 )~.
\end{align}
{The sum above is over an infinite number of modular images and it is a priori unclear whether the result is convergent. The convergence depends on the detailed structure of the preamplitude $\tilde{Z}(\tau_1,\tau_2)$ itself. We shall return to this point below in Sec \ref{s2.4}.}
The modular sum \eqref{mod-sum-0} is similar to a Poincar\'e series and, given \eqref{bootstrap1}, it is invariant under independent modular transformations. This can be seen as follows
\begin{align}
Z(\gamma_1  \tau_1, \gamma_2 \tau_2 ) &= \sum_{\gamma\,\in\, {\rm PSL}(2,\mathbb{Z} )} \tilde{Z} (\gamma_1 \tau_1 , \gamma\gamma_2 \tau_2)  =\sum_{\gamma\,\in\, {\rm PSL}(2,\mathbb{Z} )}  \tilde{Z} (\tau_1 , \gamma_1^{-1} \gamma \gamma_2 \tau_2 ) \nonumber \\
 &=\sum_{\gamma'\,\in\, {\rm PSL}(2,\mathbb{Z} )} \tilde{Z} (\tau_1 , \gamma' \tau_2 ) 
= Z(\tau_1,\tau_2). 
\end{align}
The partition function \eqref{mod-sum-0} can also be expressed as a Fourier sum (or $q$-series) and this naturally projects states onto fixed spin sectors. The BTZ threshold $k \geq 0$ then transforms into the spinning BTZ threshold, which for spin $s$, {keeps the energy above the extremality bound} $E_s \geq 2\pi \left( |s| -\tfrac{c_{\rm curr.}}{12} \right)$ \cite{Datta:2021efl,Ghosh:2019rcj,Alday:2020qkm}.\footnote{{The shift $-c_{\rm curr.}/12$ arises from the one-loop partition function.}}

Before we carry out the procedure outlined above for irrational CFTs with $\cW_N$ symmetries, we describe some essential ingredients that will be useful.


\subsection{Some ingredients of ${\mathcal{W}}_N$ CFTs}

For $\cW_N$ CFTs, the symmetry algebra is generated by modes of the stress tensor and the conserved higher spin currents, $W_m^{(s)}$ for $2\leq s \leq N$ (see \eg \cite{Bouwknegt:1992wg,Pope:1991ig} for  reviews). 
The irrational regime corresponds to the value of the central charge being larger than the number of conserved currents, $c>N-1$. In what follows, we shall require the characters of non-vacuum primaries on the torus. These are given by
\begin{align}\label{charWN}
	\chi_k(\tau) = \frac{q^{\frac{k^2}{4}}}{\eta(\tau)^{N-1}}~, \qquad \frac{k^2}{4}= h - \frac{c-(N-1)}{24}~. 
\end{align}
Here, $\eta(\tau)$ is the Dedekind eta-function and we have used a Liouville-like parametrization for the conformal dimension. As usual, these characters contain the contribution of left/right moving descendants of the primary state, $(W_{-1}^{(s_1)})^{k_1}(W_{-2}^{(s_2)})^{k_2}\cdots\ket{h}$. The case with $N=2$ reduces to the usual Virasoro CFTs. The lightest primary states in irrational $\cW_N$ CFTs scale with the central charge; this fact was found using  unitarity constraints and modular bootstrap in  \cite{Afkhami-Jeddi:2017idc}.

For carrying out the bootstrap procedure for the wormhole partition function, we shall also need the fusion kernel $S_{kk'}$ for the following S-modular transformation
\begin{align}\label{Skk'}
	\frac{\chi_k(-1/\tau)}{(-i\tau)^{N-1\over 2}}  
	= 
	\int_0^\infty dk' S_{kk'} \chi_{k'}(\tau)~. 
\end{align}
Note that the above relation is somewhat non-standard due to the presence of additional $(-i\tau)$ factors;  the origin of these factors is the moduli space volume, $V_0$, that we will encounter momentarily. We can simplify the above relation by using explicit expressions for the characters \eqref{charWN} and the S-modular transformation of $\eta(\tau)$ 
\begin{align}\label{simp-rel}
	\frac{e^{-\frac{\pi i k^2}{2\tau}}}{(-i\tau)^{N-1}}  
	= 
	\int_0^\infty dk' S_{kk'} e^{\frac{\pi i k'^2\tau }{2}}~. 
\end{align}
Our task is to extract $S_{kk'}$. We multiply both sides by $e^{\frac{\pi i q^2\tau }{2}}$, use $\tau=x+iy$ and integrate over $x$. Let's consider the RHS first
\begin{align}
	\int_{-\infty}^\infty dx \, e^{\frac{\pi i q^2\tau }{2}} \int_0^\infty dk' S_{kk'} e^{\frac{\pi i k'^2\tau }{2}} &= 4\int_0^\infty dk' \delta(q^2-k'^2) S_{kk'} = 
	\frac{2}{q} S_{kq}~. 
\end{align}
For the LHS of \eqref{simp-rel}, we expand the exponential, $e^{-\frac{\pi i k^2}{2\tau}}$, for small $k$ and integrate over $x$ term-by-term. The final result can be resummed into a modified Bessel function:
\begin{align}
	\int_{-\infty}^\infty dx \, e^{\frac{\pi i q^2\tau }{2}} \frac{e^{-\frac{\pi i k^2}{2\tau}}}{(-i\tau)^{N-1}}   =  2\pi \left(k \over q\right)^{2-N} J_{N-2} (\pi k q)~. 
\end{align}
Comparing the last two equations, we obtain the fusion kernel to be
\begin{align}\label{hs-kernel-0}
	S_{kk'} = \pi  {k^{2-N}} k'^{N-1}J_{N-2}(\pi k k')~ . 
\end{align}
For $N=2$ this becomes $S_{kk'} = \pi k' J_{0}(\pi k k')$, in agreement with the Virasoro case considered in \cite{Cotler:2020ugk}. We remark at this point that one can parametrize the conformal dimensions of a $\cW_N$ CFT in a manner similar to Toda theories, where the momenta $\vec{k}$ of vertex operators live in the root lattice of $\mathfrak{sl}(N)$ \cite{Bouwknegt:1992wg,Pope:1991ig}. An analogous fusion kernel can be derived which would be a function of the $(N-1)$ momenta. However, we shall not require that representation for our present purposes and \eqref{hs-kernel-0} will be sufficient. 


\subsection{Evaluating the wormhole amplitude}
In this subsection we evaluate the Euclidean wormhole partition function using the modular bootstrap approach. As explained in \S\ref{sec:procedure} we start by determining the preamplitude and then sum over modular images.

\subsubsection*{Fixing the preamplitude}
In order to evaluate the preamplitude we need to determine the density of primaries, $\rho(k,\bar{k})$, by imposing the bootstrap equation \eqref{bootstrap1}. {The preamplitude is understood to have its origins via a three dimensional higher spin gravity path integral, with two torus boundaries characterized by complex structures, $\tau_1, \tau_2$. In the case of pure gravity (or Virasoro CFTs at the boundaries) the path integral is proportional to the moduli space volume form, $V_0$, arising from the constrained saddle nature of the wormhole solutions \cite{Cotler:2020lxj}. This volume form can be written down in terms of the zero modes that control various features of the wormhole geometries, including the length and boundary twists, and turns out to be $\sqrt{\im(\tau_1) \im(\tau_2 )}$ \cite{Cotler:2020hgz}. } For $\cW_N$ theories at the wormhole boundaries, the current algebra OPEs can be realized via the Miura construction which involves $(N-1)$ free bosons \cite{Bouwknegt:1992wg,Pope:1991ig}. This is a generalization of the single linear dilaton realizing the Virasoro algebra. It is then reasonable to expect that the net contribution of the zero-modes is
\begin{align}
	V_0 &= \left( \im(\tau_1) \im(\tau_2 ) \right)^{\tfrac{N-1}{2}}.\label{V0}
\end{align}
A first principle derivation of this will follow from analyzing the moduli space field ranges of pure higher spin theory on $\mathbb{T}^2 \times I$. In the pure gravity case, this volume specifically arises from the symplectic measure induced by the Chern-Simons term which involves only a single time derivative. In the $SL(N,\mathbb{R})$ case, this will involve $(N-1)$ fields corresponding to the Cartan directions, which should give the factor in \eqref{V0}.

Let us now consider the preamplitude. We can check that the ansatz \eqref{ansatz} by construction is invariant under simultaneous $T$ and  $T^{-1}$ modular transformations. Therefore, the non-trivial bootstrap constraints arise only for $\tau \rightarrow -1/\tau$, and its inverse. Explicitly, it enforces the following condition
\begin{align}
	\int \frac{dk d\bar{k}}{\left( | \tau_1 |^2\right)^{\tfrac{N-1}{2} } \left( | \tau_2 |^2\right)^{\tfrac{N-1}{2} }  } \, \chi_k \left(-\tfrac{1}{\tau_1} \right) \chi_k \left(-\tfrac{1}{\tau_2} \right)\bar{ \chi}_{\bar k} &\left(-\tfrac{1}{\bar{\tau}_1} \right) \bar{\chi}_{\bar{k}} \left(-\tfrac{1}{\bar{\tau}_2} \right) \rho(k,\bar{k} ) \\ &= \int dk d\bar{k} \,\, \chi_k \left({\tau_1} \right)
	\chi_k \left({\tau_2} \right)\bar{ \chi}_{\bar k} \left(\bar{\tau}_1 \right) \bar{\chi}_{\bar{k}} \left(\bar{\tau}_2 \right) \rho(k,\bar{k} ) ,\nn 
\end{align}
where we also used, $\text{Im}\left( - {1}/{\tau} \right) =  { \text{Im}\left( \tau \right) }/{| \tau|^2 }$. Next, upon using \eqref{Skk'} in the LHS we arrive at
\begin{align}
	\int dk d\bar{k} \, S_{kk'} S_{kk''} S_{\bar k \bar{k}'} S_{\bar k \bar{k}''}  \rho(k,\bar{k} ) &= \delta(k'-k'')\delta( \bar{k}' - \bar{k}'' )\rho(k',\bar{k}')~.
\end{align}
Using the explicit form of the fusion kernel \eqref{hs-kernel-0} we obtain
\begin{align}
	\int dkd\bar{k}\,\, \left( k \bar{k} \right)^{4-2N} \rho(k,\bar{k} ) J_{N-2}(\pi kk') J_{N-2}(\pi kk'') J_{N-2}(\pi \bar k \bar k') &J_{N-2}(\pi \bar k \bar k'')\\
	&= \frac{ \delta(k' - k'') \delta( \bar k' -\bar k'')}{ \pi^4\left( k' \bar k' k'' \bar k'' \right)^{N-1} }\rho(k',\bar k' ).\nn 
\end{align} 
We multiply both sides by $k'' J_{N-2} ( \pi k'' q')$ and  $\bar k'' J_{N-2} ( \pi \bar k'' \bar q')$. Then using the orthogonality relationship of Bessel functions (see eq.~\eqref{Bessel-ortho})  we integrate over $k'', \bar k''$ and that  results in
\begin{align} 
	\frac{ \rho( q',\bar q')}{\,\,\,\,\left( q' \bar{q}'\right)^{2N-3} }&=   \frac{ \rho( k',\bar k')}{\,\,\,\,\left( k' \bar{k}'\right)^{2N-3} }~.
\end{align}
This equality can only be satisfied if the above ratio is a constant. Therefore
\begin{align}
	\rho(k,\bar k ) &= \frac{\pi^{2N-2}}{4^{N-2}\Gamma(N-1)^2}\,{\cal{C}} \left( k \bar k \right)^{2N-3}.
\end{align}
We have chosen a convenient normalization constant ${\cal C}$, such that simplifications occur later.\footnote{We are working with the convention ${\cal C} = (2\pi^2)^{-1}$ for $N=2$; this is slightly different from ${\cal C}=1$ of \cite{Cotler:2020ugk}.} Plugging this back into \eqref{ansatz} and explicitly evaluate the preamplitude to be
\begin{align}\label{pre-amp}
	\tilde{Z}(\tau_1, \tau_2 ) &= {\cal{C}}  \bigg[ Z_0(\tau_1)Z_0(\tau_2) \frac{ \left( \text{Im} (\tau_1) \text{Im}(\tau_2) \right)}{|\tau_1 + \tau_2 |^2 } \bigg]^{N-1}, \,\,\,\,\, Z_0(\tau) = \frac{1}{\sqrt{\im(\tau) }|\eta(\tau)|^2}~. 
\end{align} 
Therefore, upto the overall constant, the higher-spin preamplitude is simply  the pure gravity result raised to the $(N-1)$'th power. 
We now turn to the sum over modular images of this quantity.

\subsubsection*{Modular sum}
We are now in a position to use the preamplitude \eqref{pre-amp} to obtain the full partition function. As outlined earlier, we need to perform a sum over one-sided modular images of the preamplitude $\tilde{Z}$. As the $Z_0(\tau)$'s in \eqref{pre-amp} are modular invariant, the final result for the higher spin Euclidean wormhole partition function is 
\begin{align}\label{Z-0}
	Z_{\rm hs}(\tau_1,\tau_2) &= {\cal C}\, Z_0(\tau_1)^{N-1} Z_0(\tau_2)^{N-1} \sum_{\gamma \in{\rm  PSL}(2,\mathbb{Z}) }   \left( \frac{ \left( \text{Im} (\tau_1) \text{Im}(\gamma \tau_2) \right)}{|\tau_1 + \gamma \tau_2 |^2 }  \right)^{N-1}, 
\end{align}
with $Z_0(\tau)$ defined in \eqref{pre-amp}. 
This equation is one of the key results of this paper. Note that the wormhole amplitude above does not depend on the central charge. 
Let us now focus on the sum over modular images
\begin{align}\label{R-0}
	{\cal R} (\tau_1, \tau_2) &= \sum_{\gamma \in {\rm PSL}(2,\mathbb{Z}) }   \left( \frac{ \left( \text{Im} (\tau_1) \text{Im}(\gamma \tau_2) \right)}{|\tau_1 + \gamma \tau_2 |^2 }  \right)^{N-1}.
\end{align}
As alluded to earlier, this Poincar\'e sum will be performed for fixed spin sectors, \ie we will decompose the wormhole partition function as follows
\begin{align}\label{Zs}
	Z(\tau_1,\tau_2) &= \sum_{s_1,s_2} Z_{s_1,s_2}(\tau_1,\tau_2)~,
\end{align}
where, $s_1$ and $s_2$ denote the specific spin configuration. To this end, we rewrite $\tau_1 = z_1 + i z_2$, and, $\tau_2 = w_1 + i w_2$. The integer valued spins, $s_1$ and $s_2$,  arise as Fourier conjugate variables to $z_1$ and $w_1$ respectively. Since ${\cal R}$ is invariant under independent modular transformations (and especially the $T$ transformation) on either of the moduli, the Fourier series exists.   More explicitly this is 
\begin{align}
	{\cal R} (\tau_1, \tau_2) &= \sum_{s_1,s_2=-\infty}^\infty e^{-2\pi i z_1 s_1}e^{-2\pi i w_1 s_2} \tilde{\cal R}_{s_1,s_2}(z_2,w_2).
\end{align}
Therefore the fixed spin contribution to the wormhole partition function is given by
\begin{align}\label{Zs1s2x}
	Z_{s_1,s_2} (\tau_1, \tau_2) &= {\cal C}\, Z_0(\tau_1)^{N-1}Z_0(\tau_2)^{N-1} e^{-2\pi i z_1 s_1}e^{-2\pi i w_1 s_2}\, \tilde{\cal R}_{s_1,s_2}(z_2,w_2)\, .
\end{align}
The principal object here, $\tilde{\cal R}_{s_1,s_2}$, can be expressed using the inverse Fourier relation
\begin{align}
	\tilde{\cal R}_{s_1,s_2}(z_2,w_2)&=\int_0^1 dz_1 \int_0^1 dw_1 \, e^{2\pi  i (z_1 s_1+w_1 s_2)}\sum_{\gamma \in {\rm PSL}(2,\mathbb{Z}) }   \left( \frac{ \left( \text{Im} (\tau_1) \text{Im}(\gamma \tau_2) \right)}{|\tau_1 + \gamma \tau_2 |^2 }  \right)^{N-1} .
\end{align} 
In order to proceed, we shall exchange the order of the Fourier integrals and the modular sum. However, before we perform the Fourier integrals, it is useful to separate out the PSL$(2,\mathbb{Z})$ summation into a part which involves only the $T$ transformations and a part that involves at least a single $S$ modular transformation
\begin{align}
	\tilde{\mathcal{R}}_{s_1,s_2} = 	\mathcal{T}_{s_1,s_2} + 	\mathcal{S}_{s_1,s_2} ~. 
\end{align}
$\mathcal{T}_{s_1,s_2}$ involves a single sum over integers, while $\mathcal{S}_{s_1,s_2}$, which is more difficult to evaluate, involves summing over co-primes. 
Fortunately the universal low energy contribution to $Z(\tau_1,\tau_2)$ comes only from the part involving $T$ transformations alone and we present it here
\begin{align}
\!\!\!\!	{\cal T}_{s_1,s_2}(z_2,w_2) &= \sum_{n =-\infty}^\infty \int_0^1 dz_1 \int_0^1 dw_1\,\, e^{2\pi i z_1 s_1 + 2\pi i w_1 s_2}  \left( \frac{ \left( z_2 w_2 \right)}{(z_1 + w_1 +n)^2 + (z_2 + w_2)^2 }  \right)^{N-1}\! \label{T-0}
\end{align}
For the first integral, we may join all the summations by changing variables inside the integral of each summand, $w_1 \rightarrow w_1 + n $. We thereby get rid of $n$ dependence inside the integrand at the expense of an extended contour of integration, $(-\infty,\infty)$.  Therefore we evaluate
\begin{align}
	{\cal T}_{s_1,s_2}(z_2, w_2) &=  \int_0^1 dz_1 \int_{-\infty}^{\infty} dw_1\,\, e^{2\pi i z_1 s_1 + 2\pi i w_1 s_2}  \left( \frac{ \left( z_2 w_2 \right)}{(z_1 + w_1)^2 + (z_2 + w_2)^2 }  \right)^{N-1}.
\end{align}
Taking advantage of the infinite $w_1$ range, the integrals can be decoupled by changing variables $w= z_1 + w_1$, which results in
\begin{align}
	{\cal T}_{s_1,s_2}(z_2, w_2) &=  \left( z_2 w_2 \right)^{N-1} \int_0^1 dz_1\,\,  e^{2\pi i z_1 (s_1-s_2)} \int_{-\infty}^\infty\,dw\,\,  \frac{ e^{ 2\pi i w s_2} }{\left(  w^2 + (z_2 + w_2)^2\right)^{N-1} }  ~,
\end{align}
The $z_1$ integral results in a Kronecker delta that enforces $s_1=s_2$, whereas the $w$ integral furnishes an integral representation of the modified Bessel $K$ function. We finally get
\begin{align}\label{Ts1s2}
	{\cal T}_{s_1,s_2}(z_2, w_2) &=  \frac{2 \,(\pi z_2 w_2)^{N-1}}{\Gamma(N-1)}  \left( \frac{z_2+w_2}{ |s_2| }\right)^{\tfrac{3}{2} - N} K_{\tfrac{3}{2} - N} \left( 2\pi |s_2| ( z_2 + w_2 ) \right)\delta_{s_1,s_2}.
\end{align}
We note that when $N =2$ the above reduces precisely to the gravity answer. Details about the $S$ transformation part, which we denote with ${\cal S}_{s_1,s_2}$ can be found in Appendix \ref{app:Ss1s2}. We see, from \eqref{Ss1s2}, for the case of $N=3$, while the exponential suppression in ${\cal S}_{s_1,s_2}$ at low temperatures is the same as in ${\cal T}_{s_1 , s_2}$, the former is polynomially  suppressed while the latter is polynomially enhanced. Therefore, as stated before, we can see that low temperature behaviour is dominated by ${\cal T}_{s_1 , s_2}$. Putting everything together, the fixed spin contribution to the wormhole partition function takes the form
\begin{align}\label{Zs1s2}
	Z_{s_1,s_2} (\tau_1,\tau_2) &= {\cal C} \bigg(Z_0(\tau_1) Z_0(\tau_2)\bigg)^{N-1} e^{-2\pi i (\text{Re} (\tau_1)s_1 + \text{Re} (\tau_2) s_2)}\left[ {\cal T}_{s_1,s_2} + {\cal S}_{s_1,s_2}  \right].
\end{align}

\subsection{Comparison to other ensembles}\label{s2.4}

Now that we have obtained the partition function of the Euclidean wormhole in higher spin gravity \eqref{Z-0}, it is useful to contrast the result with other ensembles considered in the recent past. 

Let us first consider wormholes in pure gravity, which correspond to irrational CFTs with Virasoro symmetry at the boundaries. The result for the wormhole amplitude was derived in \cite{Cotler:2020hgz,Cotler:2020ugk} and is given by
\begin{align}\label{pure-grav}
	Z_{\rm pure\,grav.}(\tau_1,\tau_2)&=\frac{1}{2\pi^2}Z_0(\tau_1,\btau_1) Z_0(\tau_2,\btau_2)  \sum_{\gamma \in \mathrm{PSL}(2,Z) }
	\frac{\im(\tau_1)\im(\gamma\tau_2)}{|\tau_1+\gamma \tau_2|^2}~. 
\end{align}
The higher spin result \eqref{Z-0} along with \eqref{R-0}, agrees with the above upon setting $N=2$. The overall constant $1/2\pi^2$ is fixed using the JT gravity limit; unfortunately, we do not have an analogous 2d higher-spin gravity computation at our disposal to fix the overall constant $\mathcal{C}$ in \eqref{Z-0}. Note that the Poincar\'e series in \eqref{pure-grav} does not converge and needs to be suitably regulated \cite{Cotler:2020hgz}. The higher spin amplitude, on the other hand, has convergence built in. In fact, the `zeta-function' used in \cite{Cotler:2020hgz} to regularize the Poincar\'e sum of \eqref{pure-grav} is exactly the same as the one appearing in the higher spin wormhole partition function \eqref{R-0}.\footnote{{Please refer of \cite[eq. (4.2) and (4.3)]{Cotler:2020ugk}. As a function of $N$, the Poincar\'e sum  in \eqref{R-0}  has a simple pole only at $N \to 2^+$.}}

A more interesting comparison is with the wormhole partition function in `perturbative'  $U(1)^D \times U(1)^D$ Chern-Simons theory. This theory is the bulk dual of $D$ free bosons averaged over the Narain moduli space \cite{Maloney:2020nni,Afkhami-Jeddi:2020ezh}. The wormhole amplitude can be obtained from the connected piece of the averaged genus-2 partition function, with a diagonal period matrix, $\Omega=\text{Diag}(\tau_1,\tau_2)$. The result is  \cite{Maloney:2020nni}
\begin{align}\label{Narain}
	Z_{\rm Narain}(\tau_1,\tau_2)&=	 Z_0(\tau_1,\btau_1)^{D}Z_0(\tau_2,\btau_2)^{D}  \sum_{\gamma \in \mathrm{PSL}(2,Z)}
	\left[ 
	\frac{\im(\tau_1)\im(\gamma\tau_2)}{|\tau_1+
		\gamma \tau_2|^2}
	\right]^{D\over 2}. 
\end{align}
This amplitude is also the result of the modular bootstrap problem for $U(1)^D \times U(1)^D$  chiral algebra on the boundaries, along with the moduli space volume set as $V_0=1$ \cite{Cotler:2020ugk}. 
The $Z_0$ prefactors in \eqref{Narain}, which count the zero modes and descendant states, agree with \eqref{Z-0} upon setting $D=N-1$. This fact is isn't a mere coincidence since $N-1$ free bosons form a realization of the $\cW_N$ algebra.This realization is the Miura transformation and it works at arbitrary central charge \cite{Pope:1991ig}.\footnote{Furthermore, the non-vacuum $\cW_N$ characters \eqref{charWN} of $c>N$ theories are the as same as those of $U(1)^{N-1}$ characters. We thank Tom Hartman for this observation.}  Curiously enough, the Poincar\'e sum in \eqref{Narain} is also of the 
same form appearing in the higher spin amplitude \eqref{R-0}. However, for $D=N-1$, each term of the above sum  is the square-root of the one appearing in the higher spin amplitude \eqref{R-0}. 
For the Poincar\'e sum of the Narain ensemble \eqref{Narain}, we can use $D/2=N-1$ in \eqref{Ts1s2}. to get the fixed spin sector contribution
\begin{align}\label{T-Narain}
	\mathcal{T}_{s_1,s_2}=  \frac{ 2{\pi^{D/2} } }{\Gamma (D/2)} (z_2 w_2)^{D/2} \left(\frac{z_2+w_2}{s_2 }\right)^{\frac{1-D}{2}} K_{\frac{1-D}{2}}(2\pi s_2(z_2+w_2) )\,  \delta_{s_1,s_2}~. 
\end{align}
For $z_2=\beta_1$ and $w_2=\beta_2$ the result agrees with \cite[first line of (3.20)]{Cotler:2020hgz}. 
The factor $(z_2 w_2)^{D/2}$ cancels out exactly with factors from $Z_0(\tau_1,\btau_1)^{D}Z_0(\tau_2,\btau_2)^{D}$; this leads to an absence of a ramp in the spectral form factor. 

Despite the minor differences in the wormhole amplitude, the Narain and higher spin ensembles have very different energy eigenvalue statistics. To analyze this in detail, it is beneficial to extract the spectral density correlations from $Z_{\rm hs}(\tau_1,\tau_2)$. This is the topic of the next section.


\section{Spectral statistics from the wormhole amplitude}
\label{sec:spectral-studies}
In a quantum chaotic system, it is universally expected that there is repulsion amongst  energy eigenvalues \cite{PhysRevLett.52.1, stockmann_1999}. It has recently emerged that the Euclidean wormhole encodes analogous features for black hole microstates. The objective of this section is to quantify these features for the higher spin wormhole and understand the details of the dual ensemble description. We shall do this by studying the spectral form factor and the spectral density 2-point function.

\subsection{The spectral form factor}\label{sec:sff}
{Random matrix theory captures very universal features of quantum chaotic systems.} In this context, 
the spectral form factor (SFF) serves as a  useful tool towards diagnosing quantum chaos.  
The SFF is defined as follows 
\begin{align}\label{sff-def}
	g(\beta,t) = \vev{Z(\beta_1)Z(\beta_2)} = \vev{Z(\beta+it)Z(\beta-it)}~. 
\end{align}
{ The factors, $Z(\beta_1)$ and  $Z(\beta_2)$, denote the partition functions with inverse temperatures $\beta_1$ and $\beta_2$ which are analytically continued to $\beta + i t$ and $\beta-it$. The SFF is a simpler proxy for Lorentzian two-point correlation functions, \ie the SFF depends only on the details of spectrum and has the dependence on matrix elements of operators stripped off. The product $Z(\beta+it)Z(\beta-it)$ probes the discreteness of the spectrum,  and this aspect is realized by considering the late time behaviour of the quantity  \cite{Cotler:2016fpe}. Furthermore, in chaotic systems the SFF exhibits an universal profile \cite{Haake1991} -- this consists of an initial dip, followed by a linear ramp and then a plateau at very late times. In the above definition the average is taken over ensemble irrational CFTs with $\cW_N$ symmetries.\footnote{{The details of the averaging of this is unknown at the moment. However, we imagine the wormhole amplitude can be reproduced by a suitable averaging over the moduli of the conformal manifold, along the lines of what has been done for free theories \cite{Maloney:2020nni,Afkhami-Jeddi:2020ezh}}.} On the other hand, in random matrix theory, an integral over matrices plays the role of averaging. The linear ramp in particular, is related to the spectral rigidity of random matrix eigenvalues. %
The averaging operation also leads to a loss of factorization. In our case the non-factorization is geometrically realized via the three dimensional wormhole geometries, and the connected piece of the 2-point  function \eqref{sff-def} is built into the bootstrapped wormhole amplitude \eqref{Z-0}. 
}

In order to extract the SFF however, one needs to focus on a superselection sector of the wormhole partition function.\footnote{We thank Kristan Jensen for emphasizing this point.} In our context this corresponds to first focusing on the contribution of primaries  of $Z(\tau_1,\tau_2)$ within a fixed spin-sector. We can obtain this directly by stripping off the descendant counting functions from the spin-decomposed amplitude ${Z}_{s_1,s_2}(\tau_1,\tau_2)$ in equation \eqref{Zs1s2} 
\begin{align}
\vev{\tr_{(s_1)}^P[e^{-(\beta+it)H}]\,\tr_{(s_2)}^P[e^{-(\beta-it)H}]} &= {\cal C} \left( \text{Im} \tau_1\,\, \text{Im} \tau_2 \right)^{\frac{1-N}{2} } e^{-2\pi i (\text{Re} (\tau_1)s_1 + \text{Re} (\tau_2) s_2)}\left[ {\cal T}_{s_1,s_2} + {\cal S}_{s_1,s_2}  \right].
\end{align}
The superscript $P$ indicates the contribution from primary states.
The SFF can now be obtained by analytic continuation,
\begin{align}\label{SFF-def}
	g_{s_1,s_2}(\beta,t) = \vev{\tr_{(s_1)}^P[e^{-(\beta+it)H}]\,\tr_{(s_2)}^P[e^{-(\beta-it)H}]}~. 
\end{align}
This object encodes correlations of energy eigenvalues across sectors of fixed spin. We focus on the low-temperature regime. This corresponds to the region in parameter space where energies are close to the threshold energy of the spin-$s$ BTZ black hole 
\begin{align}
 E_s = 2\pi \left(|s| - \frac{N-1}{12}\right)	~. 
\end{align}
Let us write down the Euclidean result first. We take the boundary tori to be rectangular and set the modular parameters as $\tau_1=i\beta_1$ and $\tau_2=i\beta_2$. At low-temperatures $(\beta_{1,2}\to \infty)$ the dominant contribution arises from the sum over $T$-modular transformations. Taking into account the temperature dependence from the $Z_0$ prefactors (sans the descendant contributions) and using large argument approximation of the Bessel function \eqref{bessel-large}, we have the following result 
\begin{align}\label{low-temp-hs}
	\vev{\tr^P_{(s_1)}[e^{-\beta_1H}]\,\tr^P_{(s_2)}[e^{-\beta_2H}]}_{\rm hs} &\simeq  \frac{\pi^{N-1}\mathcal{C} }{\Gamma(N-1)}\, e^{-(\beta_1 +\beta_2) E_{s_2}}\left[ \frac{\sqrt{\beta_1\beta_2}}{\beta_1+\beta_2} \right]^{N-1} |s_2|^{N-2}\delta_{s_1,s_2}~. 
\end{align}
As a consistency check, this result agrees with pure gravity case for $N=2$ \cite[eq (3.27)]{Cotler:2020hgz}. 
It is now straightforward to perform the analytic continuation to obtain the SFF \eqref{SFF-def}. We obtain
\begin{align}
	g^{\rm (hs)}_{s_1,s_2}(\beta,t) \simeq  \frac{\pi^{N-1}\mathcal{C} }{\Gamma(N-1)}\, e^{-2\beta E_{s_2}}  \frac{ {(\beta^2+t^2)^{\frac{N-1}{2}}}}{(2\beta)^{N-1}}  
	 |s_2|^{N-2}\delta_{s_1,s_2}~. 
\end{align}
Therefore, at late times we have a power-law ramp $t^{N-1}$ within a given fixed  spin sector. This generalizes the pure gravity case ($N=2$) for which the ramp is linear.

It is worthwhile to compare the above results with the Narain ensemble (see \cite{CM} for an exhaustive study). The wormhole partition function at low temperatures is, from \eqref{T-Narain}
\begin{align}
	\vev{\tr^P_{(s_1)}[e^{-\beta_1H}]\,\tr^P_{(s_2)}[e^{-\beta_2H}]}_{\rm Narain} ~&\simeq  ~ \frac{\pi^{D/2}}{\Gamma(D/2)}\,  \frac{e^{-(\beta_1+\beta_2 )E_{s_2}}}{(\beta_1+\beta_2)^{D/2}} |s_2|^{D-2\over 2}\delta_{s_1,s_2}~,
\end{align}
where, $E_s = 2\pi (|s|-\frac{D}{12})$. Upon setting $\beta_{1,2}=\beta\pm it$, the above expression does not contain any time dependence and, therefore, the ramp is absent in the SFF. In a sense, this conclusion is well expected as the CFTs being averaged over are free theories and they do not exhibit chaotic properties.\footnote{Even after averaging the degrees of freedom (given by the central charge) equal the number of conserved currents. In this sense, the system is integrable.}

The above features are  for the connected piece of the SFF, which is given by Euclidean wormhole amplitude. There is also a disconnected component in the SFF, $\vev{Z(\beta+it)}\vev{Z(\beta-it)}$, which gives a decay and dictates early time behaviour. In the context of RMT, this decay is universal and depends only on the symmetry class of the model, for instance it is $t^{-3/2}$ for Gaussian ensembles, and, $t^{-1/2}$ for Wishart-Laguerre ensembles \cite{Meh2004}. Note that the SFF starts its life from $\langle Z(\beta)\rangle^2$. By the time the disconnected piece decays the rise coming from the connected piece begins to take over -- this is also predicted by RMT. This leads to a clear transition between the dip and the ramp. Writing down the connected SFF in spectral decomposition
\begin{align}
	g(\beta,t) = \int_0^\infty dE_1 \int_0^\infty dE_2 \,\vev{\rho(E_1) \rho(E_2)}\,e^{-\beta(E_1 + E_2 )} e^{-it( E_1- E_2)}~,\label{spectral}
\end{align}
we note that at late times, the randomness of the energies favor only small $E_1-E_2$ due to phase cancellations. This makes the ramp sensitive to nearest level distributions, which is encoded in the connected density-density correlator, $\vev{\rho(E_1)\rho(E_2)}$. For one-matrix models, this can be calculated in terms of the eigenvalue correlation kernel, $K(E_1,E_2)$, which can determine any arbitrary joint probability distribution \cite{Meh2004}. For large random matrices, when $E_1 - E_2$ is small, $K(E_1,E_2)$ is given by an universal function known as the sine-kernel \cite{Brezin:1993qg}. Using this sine-kernel, one can show that the connected SFF will start growing (as a ramp) at late times; an array of one-matrix model examples has been reviewed in \cite{Liu:2018hlr}. The dominant behaviour of the ramp is always linear, whilst there are non-linear corrections that become important at later times. 

The linear ramp of the SFF arises from the Fourier transform of the divergent contribution to the density-density correlator, $\vev{\rho(E_1) \rho(E_2) }\sim |E_1 - E_2|^{-2}$ \cite{Cotler:2016fpe}. Since for higher spins we find a power-law ramp, $t^{N-1}$, we expect a stronger divergence in the spectral density correlator. In the next subsection, we explore this expectation in detail.

\def\E{{\mathcal{E}}}
\subsection{Pair correlation function of  spectral densities}

We now want to extract the two-point function of spectral densities in a specific spin-sector. As in \eqref{spectral} this two-point function is related to the wormhole amplitude in the following manner:
\begin{align}
	\vev{\tr^P_{(s_1)}[e^{-\beta_1H}]\,\tr^P_{(s_2)}[e^{-\beta_2H}]} = \int_0^\infty dE_1 \int_0^\infty dE_2 \,\vev{\rho_{s_1}(E_1) \rho_{s_2}(E_2)}\, e^{-\beta_1 E_1-\beta_2 E_2}~. 
\end{align}
 The density-density correlator can either be obtained from the discontinuities of the double resolvent or  by directly double inverse Laplace transforming the wormhole partition function. In this section we use the latter method, and discuss the former in Appendix \ref{resolvent}. We focus on the low temperature regime, where we observed a power-law  ramp in the SFF. The partition function is given in \eqref{low-temp-hs}. The expression for $\vev{\rho_{s_1}(E_1) \rho_{s_2}(E_2)}$ is
\begin{align}
	\vev{\rho_{s_1}(E_1) \rho_{s_2}(E_2)} 
	= \mathcal{D}_{s_1,s_2}
	\int_{-i\infty}^{+i\infty} d\beta_2 e^{\beta_2 \E_2}
	\int_{-i\infty}^{+i\infty} d\beta_1  e^{\beta_1 \E_1} \left[\sqrt{\beta_1\beta_2} \over \beta_1+\beta_2\right]^{N-1}~. 
\end{align}
where, we have defined the following to lighten the notation
\begin{align}\label{light}
	\mathcal{D}_{s_1,s_2} = \frac{\pi^{N-1}\mathcal{C} }{\Gamma(N-1)} |s_2|^{N-2}\delta_{s_1,s_2}~, \qquad \E_i = E_i - E_{s_i}~. 
\end{align}

Let's consider the inverse Laplace transform (ILT) wrt $\beta_1$ first. This is
\begin{align}
	\mathcal{U}(\beta_2) = \int_{-i\infty}^{+i\infty} d\beta_1  e^{\beta_1 \E_1} \left[\sqrt{\beta_1\beta_2} \over \beta_1+\beta_2\right]^{N-1}~. 
\end{align}
The integral can be performed by expanding the $\E_1$ independent factor as a power series in $\beta_1/\beta_2$ and then integrating term by term. The result is
\begin{align}\label{Jb2}
	\mathcal{U}(\beta_2)
	&= \frac{\beta_2^{\frac{N-1}{2}}}{\Gamma(\tfrac{N-1}{2})} \E_1^{\frac{N-3}{2}} \sum_{n=0}^\infty \frac{(N-1)_n}{(\frac{N-1}{2})_n}\frac{(- \beta_2 \E_1)^n}{n!}  \nn \\
	&= \frac{\beta_2^{\frac{N-1}{2}}}{\Gamma(\tfrac{N-1}{2})} \E_1^{\frac{N-3}{2}} {}_1F_1\!\left(N-1,\tfrac{N-1}{2};-\beta_2 \E_1\right)~. 
\end{align}
Here, $(a)_n=\Gamma(a+n)/\Gamma(a)$, is the Pochammer function. We now need to inverse Laplace transform wrt $\beta_2$. The details turn out to be quite different depending on whether $N$ is even or odd. So let's consider these cases separately. 

\subsubsection*{Even $N$ }
The density-density correlator can be obtained from \eqref{Jb2} as
\begin{align}\label{ilt-zwei}
	\vev{\rho_{s_1}(E_1) \rho_{s_2}(E_2)} 
	=
	\mathcal{D}_{s_1,s_2}
	\int_{-i\infty}^{+i\infty} d\beta_1  e^{\beta_2 \E_2} \, \mathcal{U}(\beta_2) 
\end{align}
We write  $\mathcal{U}(\beta_2)$ as in the first line of \eqref{Jb2} and perform the ILT term-by-term, \ie the integral transform acts on powers of $\beta_2$. The basic identity we can use here is the following
\begin{align}\label{gamma-rel}
	\int_{-i\infty}^{+i\infty} d\beta_2\,  e^{\beta_2 \E_2} (\beta_2)^\nu= \frac{\E_2^{-\nu-1}}{\Gamma(-\nu)}~,
\end{align}
which makes sense only if $\nu$ isn't a positive integer. Upon resumming the integrated terms, we finally have the following result for the pair correlator
\begin{align}\label{rr-even}
		\vev{\rho_{s_1}(E_1) \rho_{s_2}(E_2)} _{N, {\rm even}}
	 =    \frac{(-1)^{\frac{N}{2}}(N-1)}{2\pi} \frac{(\E_1\E_2)^{\frac{N-3}{2}} (\E_1+\E_2)}{(\E_1-\E_2)^{N}} \mathcal{D}_{s_1,s_2} ~. 
\end{align}
In Appendix \ref{resolvent}, we verify the above formula using the method of resolvents for $N=2,4$. 
For $N=2$ we get the result (${\cal C} = (2\pi^2)^{-1}$ for $N=2$ in our conventions)
\begin{align}\label{virr}
	\rho_1(E_1,E_2)=-  \frac{1}{4\pi^2} \frac{\E_1+\E_2}{\sqrt{\E_1 \E_2}} \frac{1}{(\E_1-\E_2)^2} \delta_{s_1,s_2} ~.
\end{align}
This agrees with the pure gravity case.  We  observe that the two-point function \eqref{rr-even}  displays long-range eigenvalue attraction for odd $N/2$ and repulsion for even $N/2$. 
\subsubsection*{Odd $N$ }

For odd $N$ (or half-integer $N/2$), we can no longer use the identity \eqref{gamma-rel} to perform the second inverse Laplace transform. Instead, 
we can rewrite the hypergeometric function  appearing in \eqref{Jb2} in terms of Laguerre polynomials \cite{wolfram}. For $N=2P+1$ we have the following identity
\begin{align}
	{}_1 F_1 (2P,P,-x) = e^{-x} {}_1 F_1 (-P,P,-x) =\binom{2 P-1}{P}^{-1} e^{-x}L_P^{P-1}(x)~. 
\end{align}
Therefore
\begin{align}
	\mathcal{U}(\beta_2)= \frac{1}{\Gamma(P)} \E_1^{P-1} \binom{2 P-1}{P}^{-1} e^{-\beta_2 \E_1} \big[\beta_2^{P}L_P^{P-1}(-\beta_2 \E_1)\big]~. 
\end{align}
The inverse Laplace transform \eqref{ilt-zwei} can then be carried out by acting on each term of the quantity in square brackets above. For a given value of $P$, this is a finite number of terms. The ILT acts on integer powers of $\beta_2$ in the following manner
\begin{align}
	\int_{-i\infty}^{+i\infty} d\beta_2\,  e^{\beta_2 \E_{21}} (\beta_2)^m= \delta^{(m)}(\E_{21}), \qquad m \in \mathbb{Z^+}~,
\end{align}
with $\E_{21}=\E_2-\E_1$. 
The pair correlation function can then be written as
\begin{align}
	\vev{\rho_{s_1}(E_1) \rho_{s_2}(E_2)} _{N, {\rm odd}} 
	&=\frac{(N-2)!}{(N/2)!} \E_1^{\frac{N}{2}-1}\left[(\pd_{\E_{21}})^{\frac{N}{2}}
	L^{N/2 -1}_{N/2}(-\E_1\pd_{\E_{21}} ) \right] \delta(\E_{21}) \mathcal{D}_{s_1,s_2} ~,
\end{align}
This shows that the result is a linear combination of derivatives of the Dirac delta function. Unlike the case for even $N$ where we have long-range correlations, we see that the pair correlation function localizes at the contact-term singularities.  As examples, we have the following  for $\cW_3$ and $\cW_5$ 
\begin{align}
	\vev{\rho_{s_1}(E_1) \rho_{s_2}(E_2)} _{N=3} &= \pi^2 {\cal C} \left[ \delta'(\E_{21}) + \E_1 \delta''(\E_{21})\right] |s_2|\delta_{s_1,s_2} ~, \\
	\vev{\rho_{s_1}(E_1) \rho_{s_2}(E_2)} _{N=5}  &= \frac{\pi^4 \cal C }{6}\left[3\E_1 \left(3\delta''(\E_{21}) + 3\E_1 \delta'''(\E_{21}) + \frac{\E_1^2}{2} \delta^{(4)}(\E_{21}) \right)\right]|s_2|^3\delta_{s_1,s_2}~.
\end{align}

\subsection{Is there a matrix model description?}
We would now like to gain a better understanding of the spectral correlations. 
For simplicity, we shall focus on the even $N$ case for which we obtained the spectral density 2-point function to be \eqref{rr-even}. 
In the context of Gaussian unitary  ensembles (GUE), a similar behaviour of the pair correlation function is seen, $\vev{\rho(\lambda_1)\rho(\lambda_2)}\sim 1/(\lambda_1-\lambda_2)^2$. This arises from the 1d Coulomb repulsion   between the eigenvalues, $V(\lambda_i,\lambda_j)=-\log|\lambda_i-\lambda_j|$. 
Since we observe a slightly generalized spectral density correlator for $\cW_N$ CFTs, it is worthwhile to ask: what matrix ensembles or potentials $V(\lambda_i,\lambda_j)$ do they correspond to? In what follows, we shall consider a simplified version of this problem and glean the lessons. 

To keep the discussion self-contained, let us recall the well studied GUE case first. The matrix integral can be written in terms of the eigenvalues after diagonalizing the matrices and performing a unitary change of basis. It reads
\begin{align}\label{gue}
	Z_{\rm GUE} = \int [D\rho(\lambda)] e^{-S}, ~ S = - \frac{L^2}{2} \int d\lambda \, \rho (\lambda) + L^2 \int d\lambda_1d\lambda_2\, \rho(\lambda_1)\rho(\lambda_2) \log |\lambda_1-\lambda_2|~. 
\end{align}
Here, $L$ is the size of the matrix and $\rho(\lambda)$ is the unit normalized eigenvalue density.
The logarithimic potential is essentially the (exponentiated) Vandermode determinant that arises from the Jacobian while changing integration variables from matrix elements to eigenvalues. 

Our next step is to derive the pair correlation function using \eqref{gue} as the starting point -- cf.\,\cite{Cotler:2016fpe, altshuler}.  The quadratic fluctuations about the saddle  at large $L$ is 
\begin{align}\label{gue-pot}
	\delta S = - L^2 \int d\lambda_1d\lambda_2\, \delta\rho(\lambda_1)\delta\rho(\lambda_2) \log |\lambda_1-\lambda_2|~,
\end{align}
where, $\delta\rho(\lambda)= \rho(\lambda)-\rho_{\rm saddle}(\lambda)$. Fourier transforming these density fluctuations as $\delta\rho(\lambda)=\int\frac{du}{2\pi} e^{iu\lambda}\delta\rho(u)$ and performing the integrals, we get
\begin{align}\label{gue-fluc}
	\delta S = - \frac{L^2}{2} \int du\, \delta\rho(u)\frac{1}{|u|}\delta\rho(-u) ~.
\end{align} 
The $\lambda_{1,2}$ integrals in \eqref{gue-fluc}  are performed by changing variables, $r=\lambda_1-\lambda_2$, and then using the standard identity for the Fourier transform of $\log|r|$.\footnote{This identity is $
		\int_{-\infty}^\infty dk~ e^{ikx} \log |x| = - {\pi}/{|k|} - 2\pi \gamma \delta(k)$. Here, $\gamma$ is the Euler-Mascheroni constant. Also, strictly speaking, the $\pi/|k|$ should be understood in a regularized/principal value sense. The delta-function piece can be safely dropped, as the repulsion does not allow eigenvalues to coincide.} 
From this we can find the propagator and revert back to $\lambda$-space 
\begin{align}\label{gue-2}
	\vev{\delta\rho(\lambda_1)\delta\rho(\lambda_2) } \approx \frac{1}{4\pi^2 L^2} \int du\, e^{i(\lambda_1-\lambda_2)u} |u| = -\frac{1}{4\pi^2 L^2 (\lambda_1-\lambda_2)^2}~. 
\end{align}
The form of this 2-point function is analogous to the $N=2$ or Virasoro case \eqref{virr}, for small energy separations.

We would now like to reverse engineer the potential for eigenvalue interactions from a spectral density correlator, \eg given \eqref{gue-2}, we want to reconstruct \eqref{gue-pot}. A bare-bones version of \eqref{rr-even} that retains the dependence on eigenvalue differences and the overall sign is
\begin{align}\label{gen-pair}
\vev{\delta \rho(\lambda_1 )\delta \rho(\lambda_2 ) } &\sim \frac{(-1)^{N/2} }{(\lambda_1 - \lambda_2 )^{N} }~, 
\end{align}
where, $N$ is an even integer. For $N=2$ this reduces to the GUE case \eqref{gue-2}.
The inverse Fourier transform of $(\lambda_1 - \lambda_2)^{-N}$, with respect to conjugate variable $u$, is given by $(-1)^{N/2} |u|^{N-1}$. We can then write down the action of the density fluctuations in the Fourier space 
\begin{align}
\delta S &\sim \int du \, {\delta}\tilde\rho(u) \frac{1}{|u|^{N-1} } \delta \tilde\rho(-u) =\int du_1 \int du_2  \, {\delta}\tilde\rho(u_1) \frac{\delta(u_1 +u_2) }{|u_1|^{N-1} } \delta \tilde\rho(u_2) ~ . \label{Vs}
\end{align}
In the eigenvalue space, if the eigenvalue potential is given by, $V(\lambda_1 - \lambda_2) = V(r)$, then we  also have the following analogue of \eqref{gue-pot}
\begin{align}
\delta S &\sim  \int d\lambda_1d\lambda_2 \, {\delta}\rho(\lambda_1) V(\lambda_1 -\lambda_2) \delta \rho(\lambda_2)  =  \int dr\, d\lambda_2 \, {\delta}\rho(\lambda_2 + r) V(r) \delta \rho(\lambda_2) ~. 
\end{align}
The two equations above describe the same quantity and should be equivalent. In order to extract $V(r)$, we introduce the Fourier transforms as in the previous paragraph, $\delta\rho(\lambda_j)=\int\frac{du_j}{2\pi} e^{iu\lambda_j}\delta\rho(u_j)$. The integral over $\lambda_2$ yields a $\delta(u_1+u_2)$ factor, cf.~\eqref{Vs}, and we are left with the following equation for $V(r)$
\begin{align}
\int du  \,\,e^{i u r}\, V(r)  &= \frac{1}{|u|^{N-1} } ~.
\end{align}
Finally, we find that the potential takes the form
\begin{align}\label{gen-pot}
V(r) &=  r^{N-2} \big(a_N + b_N \log|r|\big)~. 
\end{align}
\begin{figure}[!t]
	\centering
	\includegraphics[width=1\linewidth]{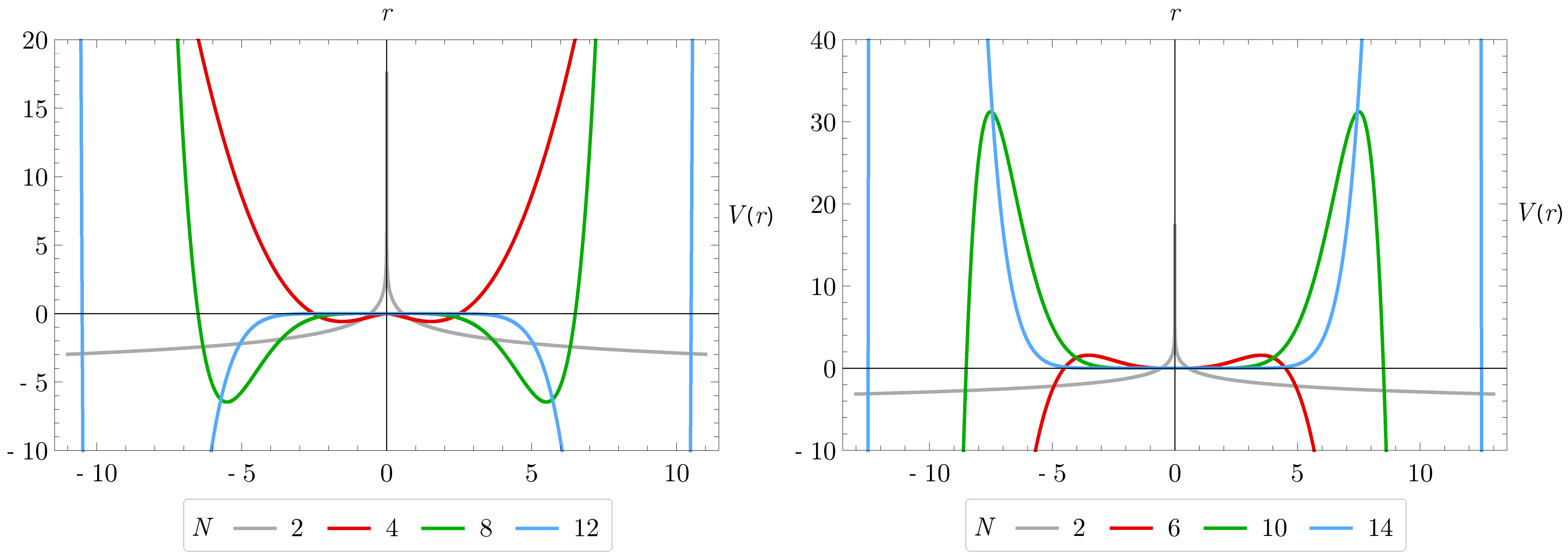}
	\caption{Plot of the potentials, $V(r)$, for $N/2$ being even (on the left) and odd integers (on the right). For $N=2$, we have $V(r)=-\log|r|$ (in gray) indicating eigenvalue repulsion in the GUE. For $N>2$ we get ``double-well" or ``double-crest'' type potentials, depending on whether $N/2$ is even or odd.}
	\label{fig:potentials}
\end{figure}
\!\!This is the generalization of the $V(r)=-\log|r|$ behaviour of the GUE case. Here $a_N$ and $b_N$ are numerical coefficients and their exact $N$-dependence isn't particularly illuminating. However, the details are such that the potentials take characteristic shapes -- see Fig.\,\ref{fig:potentials}. 
Quite strikingly, when $N/2$ is even we find a double-well shape. This is mostly attractive, with two degenerate basins located away from the origin at values that increase with $N$. This implies that most eigenvalue differences are confined in the wells and their vicinity.
On the other hand, we see that when $N/2$ is odd, the potential resembles an inverted double-well. Therefore, it is mostly repulsive with a weak attraction/confinement at very small eigenvalue differences.\footnote{A similar unexplained periodicity (in central charge) arises in the spectrum in the sphere packing/CFT$_2$ context \cite{Afkhami-Jeddi:2020hde}. It will be fascinating to uncover further connections in this direction. We thank Tom Hartman for pointing out.}

These features are in sharp contrast to the GUE matrix models. Although we considered a fairly simple form of the pair correlation function \eqref{gen-pair}, the potential \eqref{gen-pot} that gives rise to such behaviour is quite different. At this moment, it is unclear to us whether suitable matrix models (or deformations thereof) can give rise to these interactions between the eigenvalues.

\section{Discussion}
\label{sec:discuss}

In this paper we analyzed Euclidean wormholes in 3d higher spin gravity. We used modular bootstrap methods, generalizing the case of wormholes in pure gravity \cite{Cotler:2020ugk}. 
The wormhole amplitude gives the connected part of  $\vev{Z(\tau_1)Z(\tau_2)}$ of a suitably ensemble averaged, irrational, $\cW_N$ CFT. This amplitude  captures the eigenvalue statistics of black hole microstates. We observed that the spectral form factors have a power-law ramp ($\sim t^{N-1}$), and, the inferred eigenvalue dynamics exhibits many interesting features, including strong and weak attractive behaviours, as well as localization, which depend on the value of $N$. These features are novel, and perplexing at the same time -- they have not been seen previously for  pure 3d gravity \cite{Cotler:2020hgz,Cotler:2020ugk} or in the lower dimensional case of JT gravity \cite{SSS}. A drawback of our analysis is that the twist zero-mode volume, $V_0$ of equation \eqref{V0}, was argued based on symmetries of the dual CFT. It would be desirable to have a derivation of the same from the higher spin gravity perspective. 

At very late times, the spectral form factor saturates to a plateau of height $Z(2\beta)$; this follows from the very definition of the SFF and is universal. However, in order to reproduce this from the wormhole amplitude we require to take into account contributions beyond the low-temperature regime. In particular, one needs to explicitly evaluate the Kloosterman sums that include the S-transformed images of the preamplitude (see \eqref{Ss1s2} for the $N=3$ case). Reproducing the value of the plateau (which should equal the torus partition function $Z(2\tau)$) will be a good consistency check of the full bootstrapped amplitude. It is to be noted that this remains an open problem even for the pure gravity case \cite{Cotler:2020hgz}. 

Our results point towards very interesting features for the matrix ensembles that can produce the pair correlation of spectral densities. Typically random matrices exhibit eigenvalue repulsion. However, when $N$ is a multiple of 4, we found that the eigenvalues show attractive behaviour. Even though such an effective attraction is unexpected in quantum chaos discussions, there are some rare examples like \cite{Liao:2020lac}. Furthermore, some integrable systems show an exponential ramp in the SFF; an example is the $q=2$ SYK model  \cite{Winer:2020mdc,Liao:2021ofk}.
Therefore, the power-law ramp for the irrational $\cW_N$ CFTs  interpolates between the chaotic case (with a linear ramp) and the fully integrable one.
It would then be valuable to study these higher spin ensembles further, even with the goal of understanding generic eigenvalue dynamics. With this in mind and inspired by recent developments in JT gravity \cite{SSS, 	Okuyama:2019xbv, Johnson:2020heh, Johnson:2020exp,Johnson:2020mwi}, it will be fascinating to translate the analysis here in terms of a matrix model, if at all it exists.

One can imagine more general wormhole backgrounds in higher spin gravity which contain higher spin charges, in the same spirit of the higher spin black holes constructed in the past \cite{Gutperle:2011kf}. Studying the corresponding wormhole amplitude (in the grand canonical ensemble of non-zero higher spin chemical potentials) will reveal the statistics of the higher-spin charges. Unfortunately, there are technical obstacles in carrying this out; even for the case of a single torus boundary these partition functions are not known only perturbatively \cite{Kraus:2011ds,Gaberdiel:2012yb,Iles:2014gra,Apolo:2017xip,Datta:2014zpa}. Furthermore, the modular properties of the partition functions are not clearly known  which is a hindrance to the bootstrap method employed here. 

The results of this work and that of \cite{Datta:2021efl} provide information about the one- and two-point functions $\vev{Z(\tau)}$ and $\vev{Z(\tau_1)Z(\tau_2)}$ for the low-temperature regime of pure higher spin gravity in 3-dimensions. This is the near-horizon regime of near-extremal black holes in which an AdS$_2$-throat appears \cite{Ghosh:2019rcj}. For the case of higher spins, the 2d gravity description is provided by a topological BF theory \cite{Gonzalez:2018enk,Alkalaev:2019xuv,Alkalaev:2020kut,Alkalaev:2021zda}. It would be reassuring to derive $\vev{Z(\tau)}$, $\vev{Z(\tau_1)Z(\tau_2)}$ and higher point correlators which translate to BF theory on the disk, double-trumpet and geometries with multiple boundaries respectively. A related question is: how does topological recursion generalize for 2d BF theory? Given the topological nature of BF theory it is very likely that a recursive machinery will exist that would fruitfully allow the evaluation of partition functions of $n$-boundary wormholes in a genus expansion. Alternatively, one can hope to obtain the $n$-boundary amplitude by generalizing techniques of Liouville gravity, developed in \cite{Mertens:2020hbs}, to the case of Toda gravity. 
These amplitudes would enable the determination of higher moments of the spectral densities $\vev{\rho(E_1)\rho(E_2)\rho(E_3)\cdots}$. Relatedly, it can be investigated to what extent higher moments of the spectral density are fixed/constrained by the first few moments -- this is an incarnation of the truncated moment problem.


%
%
	
%
	
	


\section*{Acknowledgements}
It is a pleasure to thank Gabor Sarosi, Alba Grassi, Kristan Jensen, Shota Komatsu and Chethan Krishnan for fruitful discussions. We also thank Tom Hartman and Kristan Jensen for valuable comments on the draft. SD thanks TIFR Mumbai, CECS Valdivia and BIMSA Beijing for an opportunity to present a part of this work. DD would like to acknowledge the support provided by the Max Planck Partner Group grant MAXPLA/PHY/2018577 and {{SERB/PHY/2020334}}. 

\appendix
\section{Bessel function identities}
We list a couple of properties of (modified) Bessel functions, $J_\nu(x)$ and $K_\nu(x)$, that have been useful in our analysis. 

\begin{enumerate}
	\item The orthogonality relation of Bessel functions of the first kind is \cite{dlmf}
	\begin{align}\label{Bessel-ortho}
		\int_0^\infty  dx \,x J_\nu (ax) J_\nu (bx)  =  \frac{\delta(a-b)}{a}~. 
	\end{align}

	\item At large arguments the modified Bessel function has the following behaviour
	\begin{align}
		K_\nu (2\pi x\to \infty) \simeq   \frac{e^{-2\pi x}}{2\sqrt{x}}~.  \label{bessel-large}
	\end{align}
	
\end{enumerate}

\section{Further details on the Poincar\'e sum}\label{app:Ss1s2}
We start with the Poincare series in the Fourier space indexed by spins
  \begin{align}
 \tilde{\cal R}_{s_1,s_2}&=\int_0^1 dz_1 \int_0^1 dw_1 \, e^{2\pi  i (z_1 s_1+w_1 s_2)}\sum_{\gamma \in {\rm PSL}(2\mathbb{Z}) }   \left( \frac{ \left( \text{Im} (\tau_1) \text{Im}(\gamma \tau_2) \right)}{|\tau_1 + \gamma \tau_2 |^2 }  \right)^{N-1} .
 \end{align} 
 Using the properties of $PSL(2,\Z)$, it can be shown that $\gamma$ can be decomposed into
 \begin{align}
 \gamma &= \{ T^n\} \,\,\,\displaystyle{\cup}\,\,\, \{ T^n 
 \gamma_{c,d}
 T^m\}, \,\,\, \gamma_{c,d}=
 \begin{pmatrix} 
 [d'^{-1}]_c & [r]_{c,d'} \\
c & d'
 \end{pmatrix}
 \end{align}
 where, $T$ generates $\tau \rightarrow \tau+1$, and all other variables are integers. The first part which involves only $T$ transformations, give rise to ${\cal T}_{s_1,s_2}$, which is written down explicitly in \eqref{T-0} and evaluated in \eqref{Ts1s2}. The second part of $\gamma$ gives rise to ${\cal S}_{s_1,s_2}$. The integer $c$ is positive, while $d'\in( \mathbb{Z}/c \mathbb{Z})^*$. This means, $d'$ is an integer such that, $1\leq d' \leq c-1$ is co-prime with respect to $c$. The entry, $[d'^{-1}]_c$ is just the inverse of $d'$ modulo $c$. The entry $ [r]_{c,d'}$ is defined via the relation, $[d'^{-1}]_c d' = 1 + c [r]_{c,d'}$. Therefore we have
\begin{align}
{\cal S}_{s_1,s_2} &= \sum_{n=-\infty}^\infty  \sum_{m=-\infty}^\infty  \sum_{c\geq1, d\in (\mathbb{Z}/c\mathbb{Z})^*} \int_{0}^1 dz_1 \int_{0}^1 dw_1\,e^{2\pi i z_1 s_1 + 2\pi i w_1 s_2}   \left( \frac{ \left( \text{Im} (\tau_1) \text{Im}(T^n \gamma_{c,d} T^m \cdot \tau_2) \right)}{|\tau_1 + T^n \gamma_{c,d} T^m \cdot \tau_2 |^2 }  \right)^{N-1}.
\end{align}
 The effect of the multiple actions by the $T$ transformations, present in ${\cal S}_{s_1,s_2}$, can be dealt with in the same way, as was done with ${\cal T}_{s_1,s_2}$. Namely, via variable redefinitions, we absorb into $z_1$ and $w_1$, these {\emph {translations}} at the expense of extending the contour of integrations to the entire real line. As a result we obtain
\begin{align}
{\cal S}_{s_1,s_2} &= \sum_{c\geq1, d\in (\mathbb{Z}/c\mathbb{Z})^*} \int_{-\infty}^\infty dz_1 \int_{-\infty}^\infty dw_1\,e^{2\pi i z_1 s_1 + 2\pi i w_1 s_2}   \left( \frac{ \left( \text{Im} (\tau_1) \text{Im}(\gamma_{c,d} \tau_2) \right)}{|\tau_1 + \gamma_{c,d} \tau_2 |^2 }  \right)^{N-1}.
\end{align}
Next we note by plugging in $\gamma_{c,d}$ explicitly that the numerator within parantheses becomes independent of the integral as well as the summation, since 
\begin{align}
\frac{ \text{Im}(z) \text{Im}(\gamma_{c,d} w) }{ | z + \gamma_{c,d} w |^2 } &= \frac{\text{Im}(z) \text{Im}( w) }{ |c w z + d' z + [d'^{-1}]_c w + [r]_{c,d'}  |^2}. 
\end{align}
Therefore we have
\begin{align}
{\cal S}_{s_1,s_2} &= (z_2 w_2)^{N-1} \sum_{c\geq1, d\in (\mathbb{Z}/c\mathbb{Z})^*}  \int_{-\infty}^\infty dz_1 \int_{-\infty}^\infty dw_1\,  \frac{e^{2\pi i z_1 s_1 + 2\pi i w_1 s_2} }{|c w z + d' z + [d'^{-1}]_c w + [r]_{c,d'}  |^{2N-2} } .
\end{align}
These integrals can be performed in the complex $z_1$ and $w_1$ planes, by closing the contours appropriately. Either of the integrals present us with two $N-1$ order poles, one in the UHP while another one is its reflection into the LHP. Therefore, we need to evaluate a $(N-2)$-th order derivative to find the residue. This complicates the expressions, thus we focus on the case with $N=3$. For notational simplicity we now denote $d'$ by $d$, $[d'^{-1}]_c$ by $a$ and $[r]_{c,d'}$ by $b$. The integrals take the form
\begin{align}
\int_{-\infty}^\infty dz_1 \int_{-\infty}^\infty dw_1\,  \frac{e^{2\pi i z_1 s_1 + 2\pi i w_1 s_2} }{|c w z + d z + a w + b  |^{4} } .
\end{align}
For the $w_1$ integral we find two poles of order two. They are symmetrically placed with respect to the real $w_1$ axis. For the case, $s_2>0$ the UHP pole contributes, whereas for $s_2<0$, the residue contribution arises just from the pole located at $$ w_* = - \frac{ b + i\, a w_2 + ( d+ i \, c w_2 ) z}{a +c \, z }.$$ The location of the pole can be shown to be in the LHP since both $w_2$ and $z_2$ are non-negative. Without loss of generality, we choose the case, wherein we close the contour via the LHP, this contribution gives
$$
\frac{\pi d^4 e^{\tfrac{2\pi s_2 ( w_2 + (c w_2 - i d)( b + d z ) )}{1 + bc + cd \,z } }}{2\left( d^2 z_2 + w_2( (1+bc+cd z_1 )^2+ c^2 d^2 z_2^2 )\right)^3} \bigg( ( 1+ bc +cd z_1 )^2(2\pi s_2 w_2 -1) + 2 d^2 \pi s_2 z_2 + c^2 d^2 z_2^2 ( 2\pi s_2 w_2 -1 ) \bigg).
$$
We see that this has a third order pole in $z_1$, once again in pair. For $s_1 >0$, we shall close the contour via the UHP, and thus pick up the pole
$$
z_* = - \frac{1+bc}{c d} + \frac{i}{c} \sqrt{ \frac{z_2}{w_2} ( 1 + c^2 w_2 z_2) }.
$$
Evaluating the integral on this residue finally yields
\begin{align}
{\cal S}_{s_1,s_2} =& \sum_{c\geq1, d\in (\mathbb{Z}/c\mathbb{Z})^*} \frac{\pi^2 \, e^{2\pi i \left( \tfrac{d}{c} s_2 + \tfrac{d^{-1}}{c} s_1 \right)}}{16 c^3 \sqrt{ (1 + c^2 w_2 z_2 )^5 w_2^5 z_2^5 }} e^{-2\pi \sqrt{\tfrac{z_2}{w_2} (1 + c^2 z_2 w_2 )}\tfrac{|s_1|}{c}  -2\pi \sqrt{\tfrac{w_2}{z_2} (1 + c^2 z_2 w_2 )}\tfrac{|s_2|}{c}}\nn \\
&\times \bigg( 4\pi^2 ( 1 + c^2 w_2 z_2 )( (s_2 w_2 + s_1 z_2)^2 - 4 s_1 s_2 (c w_2 z_2)^2 )  - c^2 w_2 z_2 (1 + 4 c^2 w_2 z_2 )\nn \\
&\qquad+ 2\pi c (1 + 4 c^2 w_2 z_2 )\sqrt{(1 + c^2 w_2 z_2)w_2 z_2} (  s_1 z_2-s_2 w_2  ) 
 \bigg).
\end{align}
Note, that the sum over $(\Z/c\Z)^*$ only concerns the oscillatory exponent, which can be represented using the Kloosterman zeta function, $S(j,J;c) =  \sum_{ d\in (\mathbb{Z}/c\mathbb{Z})^*} e^{2\pi i \left( j  {d}/{c}  + J  {d^{-1}}/{c}  \right)} = S(J,j,c)$, as: 
\begin{align}
{\cal S}_{s_1,s_2} =& \sum_{c\geq1} \frac{\pi^2 \, S(s_1,s_2,c)}{16 c^3 \sqrt{ (1 + c^2 w_2 z_2 )^5 w_2^5 z_2^5 }} e^{-2\pi \sqrt{\tfrac{z_2}{w_2} (1 + c^2 z_2 w_2 )}\tfrac{|s_1|}{c}  -2\pi \sqrt{\tfrac{w_2}{z_2} (1 + c^2 z_2 w_2 )}\tfrac{|s_2|}{c}}\nn \\
&\times \bigg( 4\pi^2 ( 1 + c^2 w_2 z_2 )( (s_2 w_2 + s_1 z_2)^2 - 4 s_1 s_2 (c w_2 z_2)^2 )  - c^2 w_2 z_2 (1 + 4 c^2 w_2 z_2 )\nn \\
&\qquad+ 2\pi c (1 + 4 c^2 w_2 z_2 )\sqrt{(1 + c^2 w_2 z_2)w_2 z_2} (  s_1 z_2-s_2 w_2  ) 
\bigg).
\label{Ss1s2}
\end{align}
It is important to note that in the low temperature regime, apart from the exponential suppression the polynomial suppression goes as $1/(\beta_2 \beta_1 )^2$. The computation for other values of $N$ can be done in a similar manner.

\section{Density correlators from the resolvent}\label{resolvent}
The resolvent is a useful quantity in the context of matrix models, whose discontinuities have information about densities and their correlators. The single resolvent has information about the density of states, while the double resolvent encapsulates the pair correlation function of spectral densities. The double resolvent is defined as
\begin{align}\label{R-def}
R(E_1,E_2) &=  \bigg\langle \tr \frac{1}{H-E_1} \tr \frac{1}{H-E_2} \bigg\rangle = \int dE dE'\,\, \frac{\rho(E,E')}{(E-E_1)(E'-E_2)}~.  
\end{align} 
We can obtain the density-density correlator from its double discontinuities in the complex $E_1, E_2$ planes 
\begin{align}
R(E_1 \pm i \epsilon ,E_2 \pm i \epsilon ) &= \int dE dE'\,  \rho(E,E') 
{\cal Q}_\pm(E-E_1) {\cal Q}_\pm(E'-E_2)~,
\nonumber \\
 {\cal Q}_\pm(E-E_i) &=\bigg[{\cal P} \left( \frac{1}{E-E_i } \right)\pm i \pi \delta(E-E_i) \bigg]~.
\end{align}
Here, ${\cal P}(x)$ denotes the principal value. It then follows that
\begin{align}
\rho(E_1, E_2 ) &= \frac{ R(++) + R(--) - R(+-) - R(-+)}{(-2\pi i )^2 }. \label{ddisc}
\end{align}
From its definition \eqref{R-def}, the double resolvent can be seen to be given by the double Legendre transform of $Z(\beta_1 ,\beta_2 ) = \vev{\tr_{(s_1)}[e^{-\beta_1H}]\,\tr_{(s_2)}[e^{-\beta_2H}]}$. Therefore 
\begin{align}
R(E_1,E_2) &= \int_0^\infty d\beta_1 \int_0^\infty d\beta_2 \, e^{\beta_1 (E_1-E_{s_1})+  \beta_2 (E_2-E_{s_2}) } Z(\beta_1,\beta_2 ) \nonumber \\
&= \mathcal{D}_{s_1,s_2}
	\int_{0}^{\infty} d\beta_2 e^{\beta_2 \E_2}
	\int_{0}^{\infty} d\beta_1  e^{\beta_1 \E_1} \left[\sqrt{\beta_1\beta_2} \over \beta_1+\beta_2\right]^{N-1}~,
\end{align}
where we have used definitions of \eqref{light}. Once again the odd and even cases of $N$ require separate treatment due to very similar reasons as in the inverse Laplace transform method. Here we present a few even cases, and show that the answers agree with the ones obtained directly using inverse Laplace transforms. This provides a useful consistency check of the results. 
\begin{itemize}
\item \textbf{$N=2$ : } In this case the Legendre transform gives  
\begin{align}
R(E_1,E_2) &= \frac{\pi}{2} \frac{1}{ \sqrt{ \E_1 \E_2 } \left( \sqrt{\E_1} + \sqrt{\E_2} \right)^2 }\mathcal{D}_{s_1,s_2}.
\end{align}
Now implementing formula \eqref{ddisc}, with the discontinuities coming from the square roots, we obtain
\begin{align}
\label{rn2}
\rho(E_1, E_2 ) &= -\frac{1}{2\pi }  \frac{\E_1+\E_2}{\sqrt{\E_1 \E_2}} \frac{1}{(\E_1-\E_2)^2} \mathcal{D}_{s_1,s_2}^{(N=2)} ~.
\end{align}

\item \textbf{$N=4$ : } In this case the resolvent is
\begin{align}
R(E_1,E_2) &= \frac{3\pi}{8} \frac{1}{  \left( \sqrt{\E_1} + \sqrt{\E_2} \right)^4 }\mathcal{D}_{s_1,s_2},
\end{align}
which results in
\begin{align}
\rho(E_1, E_2 ) &= \frac{3}{2\pi } \sqrt{\E_1 \E_2} \frac{\E_1+\E_2}{(\E_1-\E_2)^4} \mathcal{D}_{s_1,s_2}^{(N=4)} ~.
\end{align}
\end{itemize}
We see that these agree with \eqref{rr-even}. 

	\begin{small}

\providecommand{\href}[2]{#2}

	\end{small}	


\end{document}